\documentclass[11pt]{article}
\usepackage[margin=1in]{geometry}
\usepackage[T1]{fontenc}
\usepackage{lmodern}
\usepackage{microtype}
\usepackage{amsmath,amssymb,stmaryrd}
\usepackage{booktabs,tabularx,multirow,array}
\usepackage{enumitem}
\usepackage[numbers,sort&compress]{natbib}
\usepackage{xcolor}
\usepackage{tikz}
\usetikzlibrary{arrows.meta,positioning,calc,fit,backgrounds,shapes.misc}
\usepackage{pgfplots}
\pgfplotsset{compat=1.18}
\usepackage{caption}
\usepackage[colorlinks=true,linkcolor=blue!50!black,citecolor=green!35!black,urlcolor=blue!50!black]{hyperref}
\usepackage{cleveref}

\setlist{itemsep=2pt,topsep=3pt,parsep=0pt}
\newcolumntype{L}[1]{>{\raggedright\arraybackslash}p{#1}}
\newcolumntype{Y}{>{\raggedright\arraybackslash}X}

\newcommand{\ob}[1]{\textsf{O#1}}
\newcommand{\rc}[1]{\textsf{RC#1}}
\newcommand{\lay}[1]{\textsf{L#1}}

\definecolor{okorange}{HTML}{E69F00}\definecolor{okblue}{HTML}{0072B2}\definecolor{okgreen}{HTML}{009E73}
\definecolor{okverm}{HTML}{D55E00}\definecolor{okpink}{HTML}{CC79A7}\definecolor{oksky}{HTML}{56B4E9}
\definecolor{okgray}{HTML}{5F6368}
\colorlet{figamber}{okorange!85!black}\colorlet{figamberfill}{okorange!14}
\colorlet{figblue}{okblue}\colorlet{figbluefill}{okblue!10}
\colorlet{figgreen}{okgreen!90!black}\colorlet{figgreenfill}{okgreen!12}
\colorlet{figgray}{okgray}\colorlet{figgrayfill}{black!5}
\colorlet{figred}{okverm}\colorlet{figredfill}{okpink!14}
\usepackage{amssymb}

\title{\bfseries Reliable Agentic Data Systems:\\ Challenges, Principles, and a Research Agenda}
\author{Xiaofei Zhang\\ University of Memphis\\ \texttt{xiaofei.zhang@memphis.edu}}
\date{November 2026}

\begin{document}
\maketitle

\begin{abstract}
Large language model (LLM) agents increasingly query, transform, remember, and act on data. Their failures are often silent: an operation executes, its output looks plausible, and the resulting claim or action is still unsupported. We argue that reliability in an agentic data system is therefore an end-to-end systems property, not merely a property of the model. Many facts needed to judge an agent's output---which state was read, whether a result is complete, what evidence supports it, and whether its inputs have changed---originate inside the data system and should not be reconstructed by the model after the fact.

We develop a reliability framework around obligations carried by agent answers and actions, identify root causes that explain why those obligations fail, and derive system guardrails that place checks where the relevant information exists. The resulting design principle is simple: record the conditions under which a result is valid, enforce what can be checked deterministically, preserve lineage and time for state that outlives a request, and expose uncertainty that cannot be discharged mechanically.

This perspective leads to a research agenda around intent, composable evidence, checkable claims, freshness of derived artifacts, transactional agent actions, and independent assurance. The goal is not an infallible agent, but a data system in which unsupported claims are difficult to release silently and the remaining uncertainty is visible to whoever must act on it.
\end{abstract}

\section{Introduction}
\label{sec:intro}

Large language model (LLM) agents are becoming a new class of data-system client. They formulate queries, call semantic operators, maintain memories and derived artifacts, and increasingly act on the results. Most data interfaces, however, were designed for a programmer or analyst who chooses an appropriate operation, interprets the result correctly, and remains accountable for the claims made from it. An agent can violate each of these assumptions while still producing an output that looks ordinary. The dangerous failures are therefore often silent: the query runs, the numbers look plausible, and the resulting claim or action is still unsupported.

\begin{figure}[t]
\centering
\resizebox{\textwidth}{!}{%
\begin{tikzpicture}[
  every node/.style={font=\small},
  box/.style={rounded corners=4pt, align=left, inner sep=5pt, line width=0.8pt},
  usr/.style={box, draw=black!55, fill=white},
  llm/.style={box, draw=figamber, fill=figamberfill},
  sys/.style={box, draw=figblue, fill=figbluefill},
  art/.style={box, draw=figgray, fill=figgrayfill},
  arr/.style={-{Stealth[length=2.2mm]}, line width=0.8pt, black!65},
  cp/.style={font=\scriptsize\sffamily\bfseries, text=white, fill=figgreen, rounded corners=2pt, inner xsep=3pt, inner ysep=1.5pt},
  ob/.style={font=\scriptsize\sffamily\bfseries, text=white, fill=figred, rounded corners=2pt, inner xsep=3pt, inner ysep=1.5pt},
  hd/.style={font=\footnotesize\bfseries\sffamily, text=black!55, anchor=west},
]
\node[usr, text width=34mm] (q) at (0,0) {``How many accounts were reachable from $a$ \emph{during March}, as recorded on March~1?''};
\node[llm, text width=30mm] (p) at (4.75,0) {\texttt{reach}($a$,\\ \quad valid: Mar 1--31,\\ \quad as of: Mar 1)};
\node[sys, text width=36mm] (e) at (9.45,0) {100 rows delivered\\[3pt]
  {\footnotesize\textbf{descriptor:} total 343 (exact),\\ delivered 100, basis Mar~1}};
\node[art, text width=52mm, anchor=west] (c1) at (12.45,1.05) {``100 accounts'' \hfill \textcolor{figred}{\textbf{$\times$ refuted}}};
\node[art, text width=52mm, anchor=west] (c2) at (12.45,0) {``343 accounts'' \hfill \textcolor{figgreen}{\textbf{$\checkmark$ certified}}};
\node[art, text width=52mm, anchor=west] (c3) at (12.45,-1.05) {``12 in Ohio'' (counted on the page)\hfill \textcolor{figgreen}{\textbf{between 12 and 255}}};
\foreach \n/\t in {q/1\; ASK, p/2\; PLAN, e/3\; EXECUTE, c1/4\; CLAIM}{\node[hd] at (\n.west |- 0,1.85) {\t};}
\draw[arr] (q) -- (p);
\draw[arr] (p) -- (e);
\draw[arr] (e.east) -- ++(0.25,0) |- (c1.west);
\draw[arr] (e.east) -- (c2.west);
\draw[arr] (e.east) -- ++(0.25,0) |- (c3.west);
\node[cp, anchor=south east] at ($(p.north east)+(-0.62,0.03)$) {L1}; \node[cp, anchor=south east] at ($(p.north east)+(0,0.03)$) {L2};
\node[cp, anchor=south east] at ($(e.north east)+(0,0.03)$) {L4};
\node[cp, anchor=south east] at ($(c1.north east)+(0,0.03)$) {L5};
\node[ob, anchor=north] at ($(q.south)+(0,-0.06)$) {O1: valid or recorded time?};
\node[ob, anchor=north] at ($(e.south)+(0,-0.06)$) {O3\; O4};
\node[ob, anchor=north east] at ($(c3.south east)+(0,-0.06)$) {O5};
\draw[-{Stealth[length=2.2mm]}, line width=1.1pt, black!40] (-1.8,-2.95) -- (17.8,-2.95);
\foreach \x/\lab in {0/{Mar 1: answer saved},6.2/{Mar 10: ingest error corrected},12.0/{Apr 2: answer reused in a fraud report}}{
  \fill[black!60] (\x,-2.95) circle (2.4pt);
  \node[font=\footnotesize, anchor=south, text=black!75] at (\x,-2.88) {\lab};
}
\node[art, text width=52mm, anchor=north west] (fr) at (12.45,-3.35) {lineage check: \textsc{possibly stale}\\ $\Rightarrow$ refresh before reuse};
\draw[arr] (12.0,-2.95) |- (fr.west);
\node[cp, anchor=south east] at ($(fr.north east)+(0,0.03)$) {L6};
\node[ob, anchor=north east] at ($(fr.south east)+(0,-0.06)$) {O7};
\node[font=\footnotesize, text=black!75, align=left, anchor=north west] at (-1.8,-3.35)
  {``What did we believe on March 1?'' remains answerable: history is kept (L0).\\[2pt]
   ``Are you sure it isn't 100?'' changes nothing without new evidence (O8).};
\end{tikzpicture}}
\caption{The running example. A question becomes a plan; the plan returns 100 rows with a descriptor recording an exact
total of 343; the agent states claims. Certification is certain-answer reasoning over the descriptor: ``100'' is
refuted, ``343'' is certified, and a count filtered over the page is certain only as an interval. When a correction
arrives later, the saved answer's lineage marks it possibly stale before it is reused. Green labels are checkpoints
(\Cref{sec:layers}); red labels are obligations (\Cref{tab:obligations}).}
\label{fig:teaser}
\end{figure}
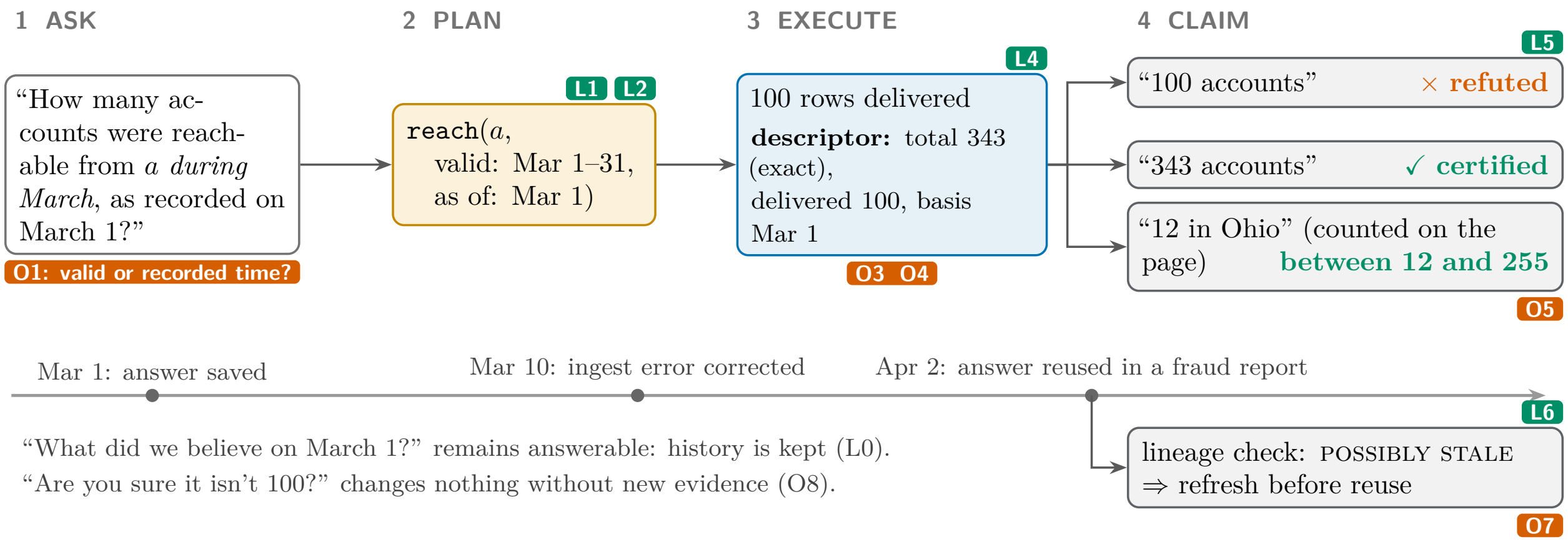

We argue that reliability in an agentic data system should be treated as an end-to-end contract between the agent and the data system, not merely as a property of the model. Better models can reduce planning and interpretation errors, but many facts needed to judge an answer are known only inside the system: which snapshot was read, whether a result is complete, which approximation was used, what evidence supports a derived value, and whether the underlying inputs have since changed. If those facts are not recorded and enforced at the interface, the agent is asked to reconstruct system state that the system itself could have made explicit. Our central principle is therefore simple: \emph{record what claims depend on, enforce what the system knows, and expose what remains unresolved} (\Cref{fig:teaser}).

This perspective matters because several failure modes that are often grouped under ``LLM accuracy'' have different system remedies. A query can execute correctly while answering the wrong interpretation of a user's request~\cite{mahmud2026exploring,li2026beyond}; a result can be locally correct but incomplete because an interface returned only a page or an approximate answer; a saved answer or memory can become invalid after the data, model, prompt, or tool changes; and an action can be unsafe even when the facts used to choose it were read correctly~\cite{li2026readingright}. These are not all model failures. Some arise because the system never recorded a condition, some because it recorded the condition but did not enforce it, and some because the remaining uncertainty genuinely belongs to the agent or user. Reliability requires making those boundaries explicit.

Database systems are well positioned to do this. Data management already provides mature ideas for transactions, integrity constraints, provenance, temporal data, access control, view maintenance, approximate query processing, and query verification. Agent engineering independently moves computation out of the model through tools, programs, solvers, and runtime checks~\cite{gao2022pal,kambhampati2024llms}. Recent data-management systems increasingly treat LLMs as persistent clients that plan, remember, and act rather than as one-shot query generators~\cite{liu2025supporting,kim2026confining}. The opportunity is to connect these lines of work around an explicit reliability contract whose enforceable parts live in the data system.

\paragraph{What this paper does.}
We first define reliability as a set of obligations carried by an agent's answers and actions, together with a minimal model of how those obligations can be discharged (\Cref{sec:model}). We then identify recurring root causes that explain why apparently plausible agent behavior becomes unreliable (\Cref{sec:rootcauses}). From those causes we derive a protection architecture that places deterministic checks, provenance, time, completeness metadata, and release gates at the points where the relevant information exists (\Cref{sec:layers}). Finally, we outline open systems problems that remain after these guardrails are in place (\Cref{sec:open}). Our goal is not to propose one complete architecture, but to make the reliability problem concrete enough that the database community can reason about what should be guaranteed, where it should be enforced, and how it should be evaluated.

\paragraph{Scope and literature grounding.}
Our focus is the boundary between an agent and the data systems it reads, transforms, and changes. We discuss model behavior when it changes what the system can safely assume, but we do not attempt a comprehensive survey of alignment, reasoning, or general agent safety. We also do not claim an exhaustive bibliometric map of agentic data systems. The literature is used selectively to establish observed failure modes, identify mechanisms that already solve part of the problem, and support the open questions developed later.

We began from recent work on LLM- and agent-driven data systems and followed citation chains into classical database mechanisms and adjacent work on verification, uncertainty, security, and ML systems. We retain references only when they support a concrete claim, provide a useful mechanism or counterexample, or clarify an open problem. Consequently, absence from the bibliography should not be read as evidence that a mechanism or system does not exist; the references are evidence for the argument, not the object of the paper.

\section{A System Model and What ``Reliable'' Means}
\label{sec:model}

\subsection{System model}

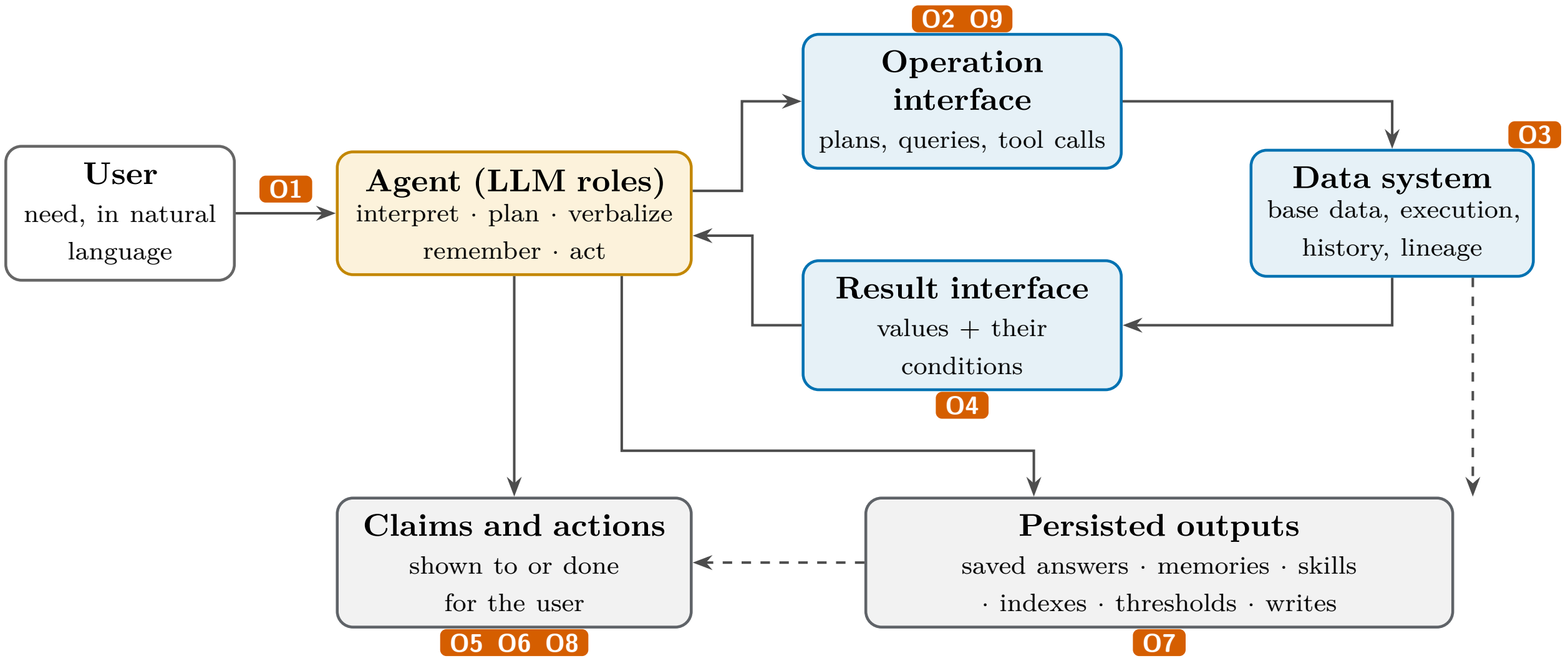
\begin{figure}[!htbp]
\centering
\resizebox{\textwidth}{!}{%
\begin{tikzpicture}[
  every node/.style={font=\small},
  box/.style={rounded corners=5pt, align=center, inner sep=5pt, line width=0.9pt, minimum height=11mm},
  llm/.style={box, draw=figamber, fill=figamberfill},
  sys/.style={box, draw=figblue, fill=figbluefill},
  art/.style={box, draw=figgray, fill=figgrayfill},
  usr/.style={box, draw=black!60, fill=white},
  arr/.style={-{Stealth[length=2.2mm]}, line width=0.8pt, black!70},
  ob/.style={font=\scriptsize\sffamily\bfseries, text=white, fill=figred, rounded corners=2pt, inner xsep=3pt, inner ysep=1.5pt},
]
\node[usr, text width=22mm] (user) at (0,0) {\textbf{User}\\{\scriptsize need, in natural language}};
\node[llm, text width=36mm] (agent) at (4.4,0) {\textbf{Agent (LLM roles)}\\{\scriptsize interpret $\cdot$ plan $\cdot$ verbalize\\ remember $\cdot$ act}};
\node[sys, text width=32mm] (opif) at (9.4,1.25) {\textbf{Operation interface}\\{\scriptsize plans, queries, tool calls}};
\node[sys, text width=32mm] (resif) at (9.4,-1.25) {\textbf{Result interface}\\{\scriptsize values + their conditions}};
\node[sys, text width=28mm] (db) at (14.2,0) {\textbf{Data system}\\{\scriptsize base data, execution,\\ history, lineage}};
\node[art, text width=36mm] (claims) at (4.4,-3.9) {\textbf{Claims and actions}\\{\scriptsize shown to or done for the user}};
\node[art, text width=62mm] (arts) at (11.6,-3.9) {\textbf{Persisted outputs}\\{\scriptsize saved answers $\cdot$ memories $\cdot$ skills $\cdot$ indexes $\cdot$ thresholds $\cdot$ writes}};
\draw[arr] (user) -- node[ob, above=3pt]{O1} (agent);
\draw[arr] ([yshift=2.5mm]agent.east) -- ++(0.55,0) |- (opif.west);
\draw[arr] (opif.east) -| (db.north);
\draw[arr] (db.south) |- (resif.east);
\draw[arr] (resif.west) -- ++(-0.55,0) |- ([yshift=-2.5mm]agent.east);
\draw[arr] (agent.south) -- (claims.north);
\draw[arr] ([xshift=12mm]agent.south) -- ++(0,-1.95) -| ([xshift=-14mm]arts.north);
\draw[arr, dashed] ([xshift=9mm]db.south) -- ([xshift=9mm]db.south |- arts.north);
\draw[arr, dashed] (arts.west) -- (claims.east);
\node[ob, anchor=south] at (opif.north) {O2\, O9};
\node[ob, anchor=north] at (resif.south) {O4};
\node[ob, anchor=south] at (db.north east) {O3};
\node[ob, anchor=north] at (claims.south) {O5\, O6\, O8};
\node[ob, anchor=north] at (arts.south) {O7};
\end{tikzpicture}}
\caption{System model. An agent interprets a need, plans operations through an operation interface, receives
results through a result interface, and emits claims, actions and persisted outputs that later work depends on.
Red labels mark where each obligation of \Cref{tab:obligations} is asserted.}
\label{fig:model}
\end{figure}

\Cref{fig:model} shows the system model used throughout the paper. A user states a need in natural language, and an
agent built around one or more LLMs may play five distinct roles: it \emph{interprets} the request, \emph{plans}
operations over a data system (for example, SQL, dataflow plans, semantic operators, or tool calls), \emph{verbalizes}
results as claims, \emph{remembers} information in saved answers, memories, or skills, and \emph{acts} by writing data,
sending messages, or triggering workflows. The data system stores base data and executes operations. The two interact
through an \emph{operation interface}, through which plans enter the system, and a \emph{result interface}, through
which values and their known conditions return. Crucially, outputs can outlive the request that produced them: answers
are saved, memories are written, skills are admitted, and later work depends on these artifacts even as the underlying
data changes.

\subsection{Obligations}

Dependability engineering distinguishes \emph{faults} (causes), \emph{errors} (deviations of internal state) and
\emph{failures} (deviations of delivered service)~\cite{avizienis2004dependable}. For an agentic data system,
correct service is not one property. We use a running example throughout.

\begin{quote}\small
\textbf{Running example.} A user asks: \emph{``How many accounts were reachable from account $a$ during March,
as recorded on March~1?''} The agent plans a temporal reachability query over valid time March~1--31 and belief
(transaction) time March~1. The operator returns its default page of 100 rows and reports a total of 343. On
March~10 an ingest error is corrected, and the saved answer is later reused in a fraud report.
\end{quote}

\noindent A claim such as ``343 accounts'' makes several assertions at once, and those assertions can fail independently. We call
them \emph{obligations} (\Cref{tab:obligations}). The first five follow the lifecycle of a single request: interpret the
need (\ob{1}), admit a valid plan (\ob{2}), execute it correctly (\ob{3}), describe the result and its conditions
(\ob{4}), and state only claims supported by that result (\ob{5}). \ob{6} concerns the policy that decides whether to
answer, ask, abstain, or escalate. \ob{7} and \ob{8} extend reliability over time and across interactions, respectively,
while \ob{9} governs side effects and the release of derived results. We keep \ob{7} and \ob{8} separate from
\ob{3}--\ob{5} because they require different state and mechanisms: history and lineage for temporal validity, and
session or memory state for interaction stability. \ob{9} is similarly meaningful only within a declared scope: a
rollback that restores the database but not the caches, files, or services an agent touched restores less than it
claims~\cite{ang2026agentic}.

\begin{table}[!htbp]
\centering\small
\caption{Obligations, instantiated on the running example.}
\label{tab:obligations}
\begin{tabularx}{\textwidth}{@{}l L{2.6cm} Y L{4.6cm}@{}}
\toprule
 & \textbf{Obligation} & \textbf{What it asserts} & \textbf{In the running example} \\
\midrule
\ob{1} & Intent fidelity & The executed operation represents the user's need (definitions, scope, windows, tie-breaks). & ``during March'' meant valid time, not when the contact was recorded \\
\ob{2} & Admissibility & The plan is well-formed, references existing objects and result fields, uses grounded identifiers, and fits its budget. & $a$ came from the resolver; the plan reads \texttt{rows\_total}, a field the operator emits \\
\ob{3} & Execution correctness & The operation was evaluated correctly over its stated basis (valid-time window, belief time, versions). & reachability computed over the March~1 belief state \\
\ob{4} & Evidence & The result's conditions are known: execution and delivery completeness, exact cardinality, approximation, provenance. & delivered 100 of 343; total exact; no approximation \\
\ob{5} & Claim faithfulness & Each stated claim is entailed by the result and its conditions. & ``343'' is entailed; ``100'' is not \\
\ob{6} & Risk-controlled selectivity & The release policy answers, asks, abstains or escalates in line with residual uncertainty. & ask what ``during March'' means if plausible interpretations produce different results \\
\ob{7} & Temporal validity & The output, and anything derived from it, remains valid as data, models and environment change. & after the March~10 correction, the saved answer is flagged before reuse \\
\ob{8} & Interaction stability & Expressed beliefs change only through evidence, across turns, sessions and memory. & ``are you sure it isn't 100?'' does not change the answer \\
\ob{9} & Effect and release safety & Writes, actions and releases of derived results are authorized, atomic or compensable, confined to a declared scope, and auditable. & the fraud report is filed once, shown only to authorized readers, with its basis attached \\
\bottomrule
\end{tabularx}
\end{table}

\subsection{A minimal formal model}
\label{sec:formal}

We make the vocabulary precise enough to state guarantees; we do not attempt a full semantics.

\paragraph{Outputs.}
An output is a tuple $o = \langle c, \pi, r, \delta, \beta, A\rangle$: typed claims $c$, the plan $\pi$ that was
executed, its result $r$, a result descriptor $\delta$ (execution and delivery completeness, exact cardinality
if known, approximation, provenance), a basis $\beta$ (valid-time window, transaction time, versions of data, model,
prompts and tools), and the interpretation assumptions $A$ the agent made. Effects are outputs whose claims are
replaced by actions.

\paragraph{Assertion rule.}
By default, an output with claims asserts \ob{1}--\ob{5} and \ob{7} for those claims; releasing it at all asserts
\ob{6} for the policy that released it; an output in a session asserts \ob{8} relative to the session's earlier
outputs; an effect asserts \ob{9}. An output \emph{withdraws} an obligation only by weakening its claim: ``at least
100'' asserts less than ``343,'' and an abstention asserts nothing about the answer. Stating conditions is not a
disclaimer; it is part of discharging \ob{4}.

\paragraph{Discharge.}
An obligation $O$ is discharged for $o$ by a certificate $\chi = \langle O, k, T, H\rangle$, where $k$ is a
guarantee kind, $T$ the trusted components it relies on, and $H$ its assumptions. We use five kinds.
\emph{By construction}: the failure cannot be expressed (a plan language without free arithmetic; identifiers only
from a resolver). \emph{Verified}: a deterministic check establishes the property, sound relative to $T$ and $H$ (a
static plan verifier; a claim checked against a descriptor; a Merkle proof of completeness). \emph{Statistical}: the
property holds with stated probability under stated distributional assumptions and a named label source, its
\emph{oracle of record} (conformal risk control; approximate answers with error bounds). \emph{Monitored}:
violations are detected after the fact at a measured rate. \emph{Human}: a person discharges it.

Certificates come at two grains. By-construction and verified certificates are \emph{per output}: they say something
about this answer. Statistical certificates are \emph{population} statements: they bound the rate of violations over a
distribution of requests (marginally, or conditionally on release), and an individual output only inherits the
statement that it was released by a policy that carries one. \ob{6} is always discharged at the population grain, for
a release policy, and \ob{8} for a session. Mixing the grains is a common source of overclaiming, and we keep them
apart below.

\paragraph{Soundness and reliability.}
An output is \emph{obligation-sound} if every obligation it asserts has a certificate whose assumptions hold, or is
surfaced to the consumer as uncertified. A \emph{failure} is an unsound output: one that behaves as if an
undischarged obligation were discharged. Soundness alone is trivial to achieve, by never answering or by surfacing
everything as uncertified, so we report reliability on a workload as three numbers rather than one: the
\emph{coverage}, the share of requests for which an output is released; for each obligation, the \emph{certified
rate}, the share of released outputs whose assertion of that obligation carries a valid certificate rather than an
``uncertified'' flag; and the \emph{unsupported-claim rate}, the share of released outputs that are unsound. A system
that answers everything with ``uncertified'' attached has full coverage, no unsupported claims and a certified rate of
zero, and the third number exposes it. Task correctness, which requires \ob{1}, is reported separately because it can
only be measured against an oracle of record. Because no mechanism inside the data system can establish \ob{1} (see
below), an output asserts \ob{1} as uncertified by default, unless a person or a clarification step discharges it. A
system's \emph{guarantee envelope} lists which obligations it discharges for which output types, with which kind, trust
base and assumptions; everything outside it is measured, not guaranteed.

\paragraph{Why certify rather than re-run.}
A consumer could try to check a claim by re-executing the plan. In agentic settings this often fails: the data may
sit behind a black-box API or a tool that cannot be replayed, the state that was read may no longer exist (no snapshot
was retained), semantic operators are not deterministic, and re-running a costly plan doubles its cost. A certificate
records, at execution time and at the one place where the facts are known, what a later checker needs; re-execution
remains the fallback where it is possible.

\paragraph{Certification as certain-answer reasoning.}
Database theory supplies the semantics for \ob{5}. Under incomplete information, a \emph{certain answer} holds in
every database consistent with what is known~\cite{imielinski1984incomplete,libkin2016sqls}; completeness statements
close an open-world result~\cite{motro1989integrity,razniewski2011completeness}. We read the descriptor and basis as such
knowledge. Let $Q$ be the executed plan, $D_\beta$ the state named by the basis $\beta$, and $\delta$ a set of facts
about $Q(D_\beta)$: the delivered rows $R \subseteq Q(D_\beta)$, and, where reported, an exact total $|Q(D_\beta)|=n$, a
lower bound, or a completeness flag. Then
\[
\mathrm{Rep}(\delta,\beta) \;=\; \{\, S \;:\; R \subseteq S,\ S \text{ satisfies every fact in } \delta \,\}
\]
is the set of results the system cannot rule out, and a claim $\varphi$ about the result is \emph{certified} iff it
holds for every $S\in\mathrm{Rep}(\delta,\beta)$. In the running example, the descriptor reports an exact total of 343
over the March~1 belief state, so ``343 reachable'' is certified and ``100 reachable'' is refuted; returned rows are
certain members; ``no account in Ohio was reachable'' is not certified, because 243 rows were not delivered. For
aggregates, certain answers are ranges rather than values, as in the range semantics of scalar aggregation over
inconsistent databases~\cite{arenas2003scalar} and attribute-level bounds in uncertainty-annotated
databases~\cite{feng2019uncertainty,feng2021efficient}: if the agent counts the rows of the page that satisfy a further
filter, say 12, the certain count over the full result is the interval $[12,\, 12+(343-100)]$, not 12; a maximum over
the page is a certain lower bound; and with no reported total, a count is only certainly at least the page count.
Claims about the world, rather than about the database, are never certified by this reasoning. The same three-valued
shape recurs in freshness: \textsc{fresh} is certain on the current state, \textsc{possibly stale} is not certain,
\textsc{undecidable} is unknown. Sound-but-incomplete approximations of certain
answers~\cite{libkin2016sqls,guagliardo2016making} are the formal ancestors of the one-sided verdicts used throughout
this paper.

\paragraph{Who can discharge what.}
The obligations differ in which component can discharge them. \ob{3} and \ob{4} concern facts that originate in the
data system during execution; a model cannot reconstruct a completeness or basis fact the system never reported,
however capable it is. \ob{2} and \ob{5} are checkable by deterministic programs given the right
records. \ob{7} needs lineage and history that only a system that recorded them can consult. \ob{1} relates a formal
operation to an informal need, and no mechanism inside the data system can establish it. \ob{6} and \ob{8} sit in
between: they concern the model's behavior, but their conditions (what evidence exists; whether a belief change is
backed by a write with a source) can be made explicit by the system. This asymmetry is the basis of the analysis
that follows.

\section{Root Causes}
\label{sec:rootcauses}

We now ask \emph{why} reliability obligations go undischarged. \Cref{tab:rootcauses} groups ten recurring causes by the kind of response they require. \emph{Design defects} (\rc{1}--\rc{3}) arise when an obligation is assigned to the wrong component, the information needed to discharge it is never recorded, or the information is available but never enforced. These defects can often be removed by redesigning the interface or execution path. \emph{Fault sources} (\rc{4}--\rc{7}) arise from ambiguous intent, change, defective inputs or infrastructure, and adversarial behavior; a system usually has to expose or contain them rather than eliminate them. \emph{Compounding} (\rc{8}) amplifies local failures across a trajectory, while \emph{assurance failures} (\rc{9}, \rc{10}) undermine the checks and measurements intended to catch the rest. An incident can involve several causes. The distinction matters because each class points to a different system intervention.

\begin{table}[!htbp]
\centering\small
\caption{Root causes, the obligations they most affect, representative evidence, and the response a data system
can give.}
\label{tab:rootcauses}
\begin{tabularx}{\textwidth}{@{}L{3.0cm} L{1.05cm} Y L{2.3cm}@{}}
\toprule
\textbf{Root cause} & \textbf{Oblig.} & \textbf{Representative evidence} & \textbf{Response} \\
\midrule
\multicolumn{4}{@{}l}{\emph{Design defects}}\\
\rc{1} Misallocation & \ob{2}--\ob{5} & wrong actions after correct reading; composition in code repairs it~\cite{li2026readingright}; invented result fields & remove: reassign \\
\rc{2} Information absence & \ob{3}, \ob{4}, \ob{7} & results without completeness or basis; reports without denominators~\cite{li2026beyondnl}; retrieval without recall; memories without time & remove: record at the source \\
\rc{3} Enforcement absence & \ob{4}, \ob{5}, \ob{8} & recorded truncation ignored without a gate; conflicting context resolved by an implicit policy~\cite{wu2024clasheval}; unverified skill admission~\cite{li2026dynamicskills} & remove: gate \\
\multicolumn{4}{@{}l}{\emph{Fault sources}}\\
\rc{4} Intent gap & \ob{1}, \ob{6} & executed but wrong~\cite{mahmud2026exploring,klopfenstein2026spotit}; silent assumptions~\cite{li2026beyond}; ambiguity~\cite{saparina2024ambrosia} & surface: clarify, record, abstain \\
\rc{5} Change and irreproducibility & \ob{3}, \ob{7} & corrected history; superseded facts served~\cite{yadav2026temporal}; model drift~\cite{chen2023chatgpt}; nondeterministic inference~\cite{atil2024non} & contain: versions, lineage, logged outputs \\
\rc{6} Defective inputs and infrastructure & \ob{3}, \ob{9} & dirty or model-extracted base data~\cite{li2023llm}; tool faults; non-idempotent retries & contain: validation, idempotency \\
\rc{7} Subversion & \ob{2}, \ob{8}, \ob{9} & indirect injection~\cite{greshake2023not}; adaptive attacks~\cite{nasr2025attacker}; poisoned memory~\cite{chen2024agentpoison,zou2024poisonedrag} & contain: confinement, write integrity \\
\multicolumn{4}{@{}l}{\emph{Amplifier}}\\
\rc{8} Compounding & all & low repeatability (pass$^k$)~\cite{yao2024bench}; belief drift under pressure~\cite{sharma2023understanding,li2026epistemic}; short reliable horizons~\cite{kwa2025measuring}; multi-turn loss~\cite{laban2025llms}; cascading steps~\cite{zhu2025where,cemri2025multi} & contain: checks at commit points \\
\multicolumn{4}{@{}l}{\emph{Assurance failures}}\\
\rc{9} Correlated checkers & \ob{5}, \ob{6} & shared judge errors~\cite{li2026jev}; confidence uncorrelated with correctness~\cite{li2026lrm}; oracle-relative guarantees~\cite{liu2026certified} & deterministic, independent checks \\
\rc{10} Measurement gap & (all) & label errors in more than half of two benchmarks~\cite{jin2026pervasive}; flawed agent benchmarks~\cite{zhu2025establishing} & measure what is claimed \\
\bottomrule
\end{tabularx}
\end{table}

\subsection{Design defects}

\paragraph{\rc{1} Misallocation: the obligation is held by a component that cannot discharge it reliably.}
Agents are routinely asked to compose facts under rules, count, do arithmetic, carry identifiers and account for
completeness. The clearest behavioral evidence we know of comes from agents outside data systems. In Li et al.'s study
of decision models in agent loops~\cite{li2026readingright}, most wrong actions follow correct reading, and the share
grows with accuracy. Errors
concentrate on the same 141 states across 14 competent systems, so they are properties of states, not noise. When
facts provably determine the action, the gap between reading and acting is about 20 points, and composing the action
in code from the model's own fact answers closes it (0.96--1.00); feeding the same answers back to the model recovers
less and breaks some states that were already correct. The study covers choice among supplied options under rule
precedence in simulated, customer-service, web and coding environments, not arithmetic, identifiers or time, and not
data agents; we take it as evidence that misallocation exists, not as a measure of how common it is in data work. The same distinction appears when planners invent result fields or violate declared preconditions. Surface heuristics can also override unstated feasibility constraints: a
distance cue outweighs the goal by 8.7 to 38 times, and enumerating preconditions recovers only 6--9
points~\cite{li2026modelsayswalk}. A data system cannot supply common-sense constraints, but it can make
\emph{declared} preconditions (types, integrity constraints, temporal validity) checkable. For values a model would
otherwise recall from its parameters there is also a statistical argument. Under the assumptions of Kalai and
Vempala~\cite{kalai2024calibrated} (a model calibrated to its training distribution, and ``arbitrary'' facts with no
learnable pattern), the rate of hallucination on such facts is bounded below by roughly the fraction of them seen once in
training; specific identifiers, dates and counts, exactly the values a database stores, are of this kind.

The principle of reallocation comes from model research: program-aided and solver-aided
reasoning~\cite{gao2022pal,ye2023satlm}, tool use~\cite{schick2023toolformer,yao2023react},
and the finding that models cannot reliably self-correct without external feedback~\cite{huang2024selfcorrect}. The
same holds for procedural knowledge: self-generated skills give no average benefit while curated ones
do~\cite{li2026skillsbench}. What the data-system perspective adds is \emph{where} to reallocate: into components that
also record the conditions under which their outputs hold. Reallocation concerns meaning, not format: constrained decoding can guarantee structural validity without guaranteeing that the resulting operation answers the task, and format restrictions can themselves impair reasoning~\cite{tam2024let,park2024grammar}.

\paragraph{\rc{2} Information absence: the fact that would decide the claim was never recorded or exposed.}
Many tool interfaces return a first page or a capped result without stating what was left out. Interface design
matters to agents~\cite{yang2024sweagent}, and a stronger model or a friendlier message repairs misreadings of an
interface, but neither can supply a completeness or basis fact that the interface omits. Retrieval over unstructured
data has the same gap in a worse form: approximate nearest-neighbor and top-$k$ retrieval rarely report recall, so an
agent cannot know what it did not see. The pattern recurs outside databases. A predictive report is meaningful only
if it carries its target, population and failure handling in the denominator; a metric over valid outputs only is the
statistical twin of a count over an incomplete page~\cite{li2026beyondnl}. A calibrated probability is defined only
relative to an event, an information set and a population~\cite{li2026calibrationrl}. Business semantics are often
absent from schemas: one study found that a short document of measure definitions raised text-to-SQL accuracy by
17--23 points~\cite{rumiantsau2026semantic} (single dataset, preprint), and failures of successfully executed SQL have
been traced to a gap between analytical concepts and what schemas expose~\cite{mahmud2026exploring}. Persistent agent
state suffers most. Memory systems typically store extracted facts without separating when something was true from
when it was learned~\cite{packer2023memgpt,chhikara2025mem0}, with recent bi-temporal
exceptions~\cite{rasmussen2025zep,wang2026less}; skill records rarely declare the tools, model and task mix they were
validated against~\cite{li2026dynamicskills}; and few pipelines record which model version produced which artifact,
although hosted models change behavior under a fixed name~\cite{chen2023chatgpt,cai2025you}.

\paragraph{\rc{3} Enforcement absence: the fact was exposed, but nothing acted on it.}
A result interface that records completeness can prevent downstream checks from treating a partial page as a complete relation. The relevant fact is
recorded in both cases; only the gate changed the verdict. Conflicts between retrieved
context and parametric knowledge are resolved by an implicit policy: models adopt incorrect retrieved content over a
correct prior more than 60\% of the time in ClashEval~\cite{wu2024clasheval}. Memory pipelines can keep a user's
misconception and drop the correction, and most memory-system errors occur after the right memory was
retrieved~\cite{li2026epistemic}. Skill libraries show the same cause at admission: in one system
in~\cite{li2026dynamicskills}, removing the verifier dropped pass rate from 71.1\% to 41.1\%.

\subsection{Fault sources}

\paragraph{\rc{4} The intent gap.}
Intent fidelity (\ob{1}) relates a formal operation to an informal need, and no mechanism inside the system can
establish it. Verifying execution establishes that a value follows from the operation that ran, not that the operation was the
right reading of the question: SQL that executes successfully still fails analytical intents in recurring
ways~\cite{mahmud2026exploring}, and bounded verification shows test-database accuracy overstating
correctness~\cite{klopfenstein2026spotit,tremante2026spotit}. Agentic text-to-SQL systems can return correct-looking results while silently making
unverified assumptions, and often abandon clarification questions they had already identified~\cite{li2026beyond}.
Ambiguity persists even when the database is visible~\cite{saparina2024ambrosia,bhaskar2023benchmarking}, and models
recognize it poorly~\cite{liu2023re}. On real agent states, what is missing is often the next sub-goal rather than a
fact: for one fast model on ALFWorld, stating the sub-goal adds 16.3 points and facts about 5~\cite{li2026readingright}.
Specifications are also discovered rather than given: evaluators cannot fix criteria before seeing
outputs~\cite{shankar2024who}. Two smaller residuals belong here: database answers are true of the database, not the
world, so ``no matching events'' is confusable with ``outside coverage''; and the system
cannot decide the truth of a user's own assertion, even when it can require a source for it. We hypothesize that once
design defects are removed, intent becomes the dominant residual. Establishing whether intent becomes the dominant residual after system-level defects are removed remains an empirical question (\Cref{sec:open}).

\paragraph{\rc{5} Change and irreproducibility.}
\ob{7} is asserted for as long as an output is relied on, and the world moves underneath it. History is corrected, and a saved answer can depend on rows it does not contain (an absence, a threshold, a count),
so the cheap rule ``did the correction touch a row of the stored result?'' is unsound. In one preprint, similarity-indexed memory could not
tell a contradicted fact from a duplicate (AUROC 0.59), and retrieval-augmented systems served superseded values
15--40\% of the time when forced to answer~\cite{yadav2026temporal}. Weight edits fail to propagate to implied
facts~\cite{cohen2023evaluating,zhong2023mquake}, and models do not know when their knowledge dates
from~\cite{cheng2024dated}. Components change too: hosted model behavior differs across snapshots~\cite{chen2023chatgpt};
thresholds and calibrators fitted on one workload drift on another~\cite{li2026jev,li2026calibrationrl}; skills go
stale when tools or backbones change~\cite{li2026dynamicskills}. Irreproducibility compounds this: temperature-zero
inference varied accuracy by up to 15\% across identical runs in one study~\cite{atil2024non}, numerical precision and
batch configuration change long reasoning chains~\cite{yuan2025understanding}, and one report traced divergence to
batch-variant kernels, with 80 distinct completions in 1,000 runs of one prompt~\cite{he2025defeating} (not peer
reviewed). The same question over the same snapshot can therefore receive different answers, so answer-level
repeatability must be engineered, not assumed~\cite{kim2026confining}, and an audit cannot rely on re-running the model.

\paragraph{\rc{6} Defective inputs and infrastructure.}
Classical data management assumes neither clean data nor reliable infrastructure, and agents inherit both problems.
Base data are dirty, duplicated or stale; text-to-SQL benchmarks built on real databases include dirty values for this
reason~\cite{li2023llm}. Increasingly, base facts are themselves extracted from text by models, so the timeline and
identities in the store are uncertain before any agent queries them. LLMs now clean, match and integrate data
competitively~\cite{narayan2022foundation,zhang2024jellyfish}, but such work reports
accuracy, not which repairs were made on what evidence; table reasoners degrade on defective tables and rarely flag
the defect~\cite{gu2025radar,wolff2025well}; and LLM checks of LLM-extracted facts over-accept~\cite{boylan2024kgvalidator}.
Merges and repairs are usually destructive writes, although they are revisions in the bi-temporal sense, and whether
data are clean enough is itself relative to the question asked~\cite{li2021cleanml}. Tools and services time out, rate-limit and fail
partially, and an agent that retries a non-idempotent call duplicates its effect. These are ordinary faults with
ordinary mechanisms (validation, entity resolution, idempotency keys, timeouts); their absence from agent stacks is a
design choice, not a law.

\paragraph{\rc{7} Subversion.}
Adversaries target the mechanisms that discharge obligations. Text retrieved by an agent can carry
instructions~\cite{greshake2023not,debenedetti2024agentdojo}; defenses trained into models are
bypassed by adaptive attacks~\cite{nasr2025attacker,zhan2025adaptive}; prompt injection reaches databases through
generated SQL~\cite{pedro2023prompt}. Knowledge bases and memories can be poisoned with a handful of
documents~\cite{zou2024poisonedrag,chen2024agentpoison}, or through ordinary queries alone~\cite{dong2025memory}. Skills
are content that agents execute, so skill-file injection is code injection~\cite{li2026dynamicskills}. Inside a verified
system, the trust base is the target: an evidence verifier is only as sound as the components that describe results to it.

\subsection{The amplifier}

\paragraph{\rc{8} Compounding over steps, turns, agents and shared state.}
Low repeatability is the starting point: on $\tau$-bench, the probability of succeeding on all eight repetitions of a
task fell below 25\% in one domain~\cite{yao2024bench}, so a long chain of such steps rarely completes cleanly. The task length at which agents succeed 80\% of the time is much shorter than at
50\%~\cite{kwa2025measuring}. Performance drops by an average of 39\% when instructions arrive over several turns,
because models commit early and do not recover~\cite{laban2025llms}. Multi-agent systems fail at inter-agent boundaries
and at verification~\cite{cemri2025multi}; early errors cascade~\cite{zhu2025where}; a single injected falsehood can
propagate through a pipeline~\cite{li2026epistemic}; and agents that share mutable state create the lost updates and
stale reads that transactions were invented to prevent~\cite{mraz2026tracking}. Across turns, sycophancy and belief drift revise expressed beliefs in response to inputs that carry no
evidence~\cite{sharma2023understanding,hong2025measuring}, and the hazard of a
first inconsistency depends on conversation history~\cite{li2026timetoinconsistency}. Epistemic robustness pairs
stability under pressure that carries no evidence with responsiveness to pressure that does~\cite{li2026epistemic}, and
the pairing is the difficulty: tuning that lowers concession to unwarranted pressure can also suppress valid corrections
(one study reports acceptance of valid corrections falling from 41.73\% to 2.17\%). A data system cannot
fix the model's behavior, but it can make the condition explicit: a belief change backed by a write with a source is
admissible, one without is not, and the gate then needs no guess about the validity of the pressure (\lay{6}). One preprint reports that
long-horizon success decays geometrically with step count rather than with context length~\cite{mittal2026fast}; if
this replicates, checks at step boundaries matter more than context engineering. In data work, curated skills fix
single-step idioms but not multi-stage pipelines in which an early failure blocks everything
downstream~\cite{li2026skillsbench}.

\subsection{Assurance failures}

\paragraph{\rc{9} Correlated and oracle-relative checkers.}
Defense in depth works when the holes in successive layers rarely line up~\cite{reason2000human}; layers that fail for
the same reasons give little more than one. LLM-based checkers often fail for the same reasons as what they check. Models from different providers make correlated errors~\cite{kim2025correlated,goel2025great}, and judges prefer their
own generations~\cite{panickssery2024llm}, so a cascade cannot repair errors they share; judges fail most on
derivations and on elaborately written wrong answers, where confident errors rose from 1.5\% to 7.9\% in one setting,
and on reference-free prose they were near chance yet confident~\cite{li2026jev}. A single token can fool some
judges~\cite{zhao2025one}, and rewards get gamed, whether they come from learned reward models~\cite{gao2022scaling} or from verifiers whose
checks are incomplete~\cite{helff2026llms}. Self-reported
confidence of reasoning models can be essentially uncorrelated with correctness ($r=0.07$ in one
study)~\cite{li2026lrm}; fine-tuned SQL verifiers fall from about 0.77--0.79 AUROC in distribution to about 0.66 on
unseen schemas~\cite{richardson2026predicts}. Statistical guarantees are relative to their label source, and their coverage degrades when the calibration labels
are noisy~\cite{einbinder2022conformal,sesia2023adaptive}. Guarantees for
semantic-operator cascades bound deviation from an LLM ``oracle,'' not from the
truth~\cite{kang2020approximate,zeighami2025cut}, and composing them across operators has only begun~\cite{sanmartino2026stretto}. On one text-to-SQL benchmark, conformal abstention certificates calibrated on execution labels at a nominal
risk of 0.10 had a realized risk of 0.17 to 0.20 against blinded expert labels, on two
checkpoints~\cite{liu2026certified}. Calibration of each step does not imply calibration of a
trajectory~\cite{lin2026uncertainty}. A key unanswered question is whether the protection layers of an agentic system fail independently enough for defense in depth to provide real assurance.

\paragraph{\rc{10} Measurement gap.}
Experts found annotation errors in 52.8\% of BIRD Mini-Dev and 62.8\% of a Spider~2.0 subset, and re-scoring on
corrected labels reordered agent rankings~\cite{jin2026pervasive} (figures differ across versions of that paper).
Execution accuracy has false positives and negatives~\cite{zhong2020semantic,renggli2025fundamental};
benchmark noise alone can change which method looks best~\cite{wretblad2024understanding}. Flawed task setups and
rewards misestimate agent performance by up to 100\% in relative terms~\cite{zhu2025establishing}, and log inspection
reveals shortcuts invisible in scores~\cite{kapoor2025holistic}. Across 18 months of model releases, accuracy gains
yielded only modest gains in consistency, robustness and predictability~\cite{rabanser2026science}.

The distinction also clarifies the transition from diagnosis to protection. Misallocated responsibilities, missing information, and absent enforcement can often be addressed directly by the data system. Intent ambiguity, change, imperfect inputs, and adversarial behavior require containment, explicit uncertainty, or escalation. Compounding and assurance failures then determine how far a local mistake can propagate and how much confidence to place in the protections themselves. The next section turns these observations into guardrails placed along the agent--system dataflow.

\section{System Guardrails}
\label{sec:layers}

Database systems became reliable on unreliable hardware, under concurrency and with untrusted clients not by
trusting any one component but by combining narrow, checkable guarantees: logging and recovery, isolation, integrity
constraints, access control, view maintenance~\cite{haerder1983principles,gupta1999maintenance}.
The end-to-end argument says where such checks belong: a property can be guaranteed only with the knowledge of the
endpoints, and intermediate checks are worth adding for containment and performance, not as
substitutes~\cite{saltzer1984endtoend}. An agentic data system adds a new unreliable component: a stochastic author of operations, memories, and claims. The same systems logic therefore calls for guardrails that place each check where the relevant information is available.

\subsection{Design principles}
\label{sec:principles}

\begin{description}[leftmargin=0pt,labelsep=0.5em]
\item[\textbf{Allocate each obligation to the component that can discharge it.}] Composition under rules,
arithmetic, identifiers, completeness accounting and time belong to deterministic operators; interpretation, plan
proposal and verbalization belong to the model.
\item[\textbf{Record at the source.}] The facts that decide whether a claim is supported (basis, belief time,
completeness, cardinality, approximation, provenance, model and environment versions) originate where the work is
done. Middleware can enforce them but cannot reconstruct what the system never reported.
\item[\textbf{Enforce; do not merely inform.}] A condition the consumer is free to ignore will be ignored. Each
recorded condition needs a gate that acts on it.
\item[\textbf{Prefer one-sided verdicts, surface the residual, and report the price.}] When a check cannot decide,
it should fail toward the safe side (\textsc{possibly stale}, uncertified, ask, abstain) and be counted. What no
check can decide, chiefly intent, should be shown rather than hidden.
\item[\textbf{Version everything derived.}] Saved answers, memories, skills, indexes, thresholds and calibrators are
views over data, models and environments; each needs a time and a lineage so its validity can be re-decided
(\rc{5}).
\item[\textbf{Keep checks independent and the trust base explicit.}] Stacked checks help only if they do not share
failure modes, and every verified guarantee is relative to the components it trusts. Check deterministically what
can be checked deterministically; name the trust base of the rest.
\item[\textbf{Speculate freely; certify at commit points.}] Agents explore by issuing many redundant, approximate
attempts~\cite{liu2025supporting}. Exploration should be cheap and isolated; contracts apply at \emph{commit points}:
where a claim is released, an effect becomes irreversible, or an artifact is persisted for later reuse. A step that
feeds a commit point is checked before it does.
\item[\textbf{Measure both sides of every trade.}] Check the checkers. Report coverage with soundness,
false-stale with false-fresh, utility with security; pre-register thresholds; fault-inject the verifiers.
\end{description}

\subsection{Checkpoints on the dataflow}
\label{sec:stack}

\begin{figure}[!htbp]
\centering
\resizebox{\textwidth}{!}{%
\begin{tikzpicture}[
  every node/.style={font=\small},
  box/.style={rounded corners=5pt, align=center, inner sep=5pt, line width=0.9pt, minimum height=11mm},
  llm/.style={box, draw=figamber, fill=figamberfill},
  sys/.style={box, draw=figblue, fill=figbluefill},
  art/.style={box, draw=figgray, fill=figgrayfill},
  usr/.style={box, draw=black!60, fill=white},
  arr/.style={-{Stealth[length=2.2mm]}, line width=0.8pt, black!70},
  cp/.style={font=\scriptsize\sffamily\bfseries, text=white, fill=figgreen, rounded corners=2pt, inner xsep=3pt, inner ysep=1.5pt},
  cpl/.style={font=\scriptsize, text=figgreen!70!black, align=center},
]
\node[usr, text width=22mm] (user) at (0,0) {\textbf{User}};
\node[llm, text width=36mm] (agent) at (4.6,0) {\textbf{Stochastic core}\\{\scriptsize LLM: interpret $\cdot$ plan\\ verbalize $\cdot$ remember $\cdot$ act}};
\node[sys, text width=30mm] (opif) at (9.6,1.3) {\textbf{Operation interface}};
\node[sys, text width=30mm] (resif) at (9.6,-1.3) {\textbf{Result interface}};
\node[sys, text width=30mm] (db) at (14.4,0) {\textbf{Execution}\\{\scriptsize deterministic operators}};
\node[art, text width=36mm, minimum height=9mm] (sub) at (14.4,-2.6) {\textbf{L0 Substrate}\\{\scriptsize history, lineage, versions}};
\node[art, text width=36mm] (claims) at (4.6,-4.1) {\textbf{Claims and effects}\\{\scriptsize effects run via L2, L3}};
\node[art, text width=40mm] (arts) at (10.4,-4.1) {\textbf{Persisted artifacts}\\{\scriptsize answers, memories, skills}};
\draw[arr] (user) -- (agent);
\draw[arr] ([yshift=2.5mm]agent.east) -- ++(0.6,0) |- (opif.west);
\draw[arr] (opif.east) -| (db.north);
\draw[arr] (db.west) -- ++(-0.55,0) |- (resif.east);
\draw[arr] (resif.west) -- ++(-0.6,0) |- ([yshift=-2.5mm]agent.east);
\draw[arr, <->] (db.south) -- (sub.north);
\draw[arr] (agent.south) -- (claims.north);
\draw[arr] ([xshift=12mm]agent.south) -- ++(0,-2.1) -| (arts.north);
\draw[arr, dashed] (sub.west) -| ([xshift=12mm]arts.north);
\node[cp] at ($(user.east)!0.5!(agent.west)+(0,0.42)$) {L7};
\node[cpl, anchor=north] at ($(user.east)!0.5!(agent.west)+(0,-0.1)$) {clarify\\ escalate};
\node[cp, anchor=south] at ($(opif.north)+(-0.75,0)$) {L1};
\node[cp, anchor=south] at ($(opif.north)+(0.05,0)$) {L2};
\node[cpl, anchor=south west] at ($(opif.north)+(0.45,0.02)$) {contract, admission};
\node[cp, anchor=south] at (db.north east) {L3};
\node[cp, anchor=north] at ($(resif.south)+(-0.4,0)$) {L4};
\node[cpl, anchor=north west] at ($(resif.south)+(0.05,-0.02)$) {evidence};
\node[cp] at ($(agent.south)+(-0.55,-1.0)$) {L5};
\node[cpl, anchor=east] at ($(agent.south)+(-0.9,-1.0)$) {claim gate,\\ selectivity};
\node[cp, anchor=north] at (arts.south) {L6};
\node[cpl, anchor=north west] at ($(arts.south)+(0.35,-0.02)$) {freshness, write admission};
\node[draw=figred, fill=figredfill, rounded corners=3pt, minimum width=158mm, minimum height=6mm, font=\scriptsize, anchor=north] at (7.2,-5.25) {\textbf{Security} across all checkpoints: confinement, information-flow control, least privilege for reads and writes, write integrity};
\node[draw=black!55, fill=white, rounded corners=3pt, minimum width=158mm, minimum height=6mm, font=\scriptsize, anchor=north] at (7.2,-5.95) {\textbf{Evaluation and observability}: traces, fault injection of verifiers, pre-registration, pass$^k$, coverage accounting};
\end{tikzpicture}}
\caption{Protection checkpoints (green) on the dataflow of \Cref{fig:model}. The substrate \lay{0} records history,
lineage and versions that the checkpoints consult. The model is a stochastic core whose improvements raise the
probability of discharge; the checkpoints bound the consequences of failure. Effects (writes, messages, workflow
triggers) are operations: they enter through the operation interface and pass admission (\lay{2}) and execution
(\lay{3}) like queries. Security and evaluation cut across all checkpoints.}
\label{fig:layers}
\end{figure}
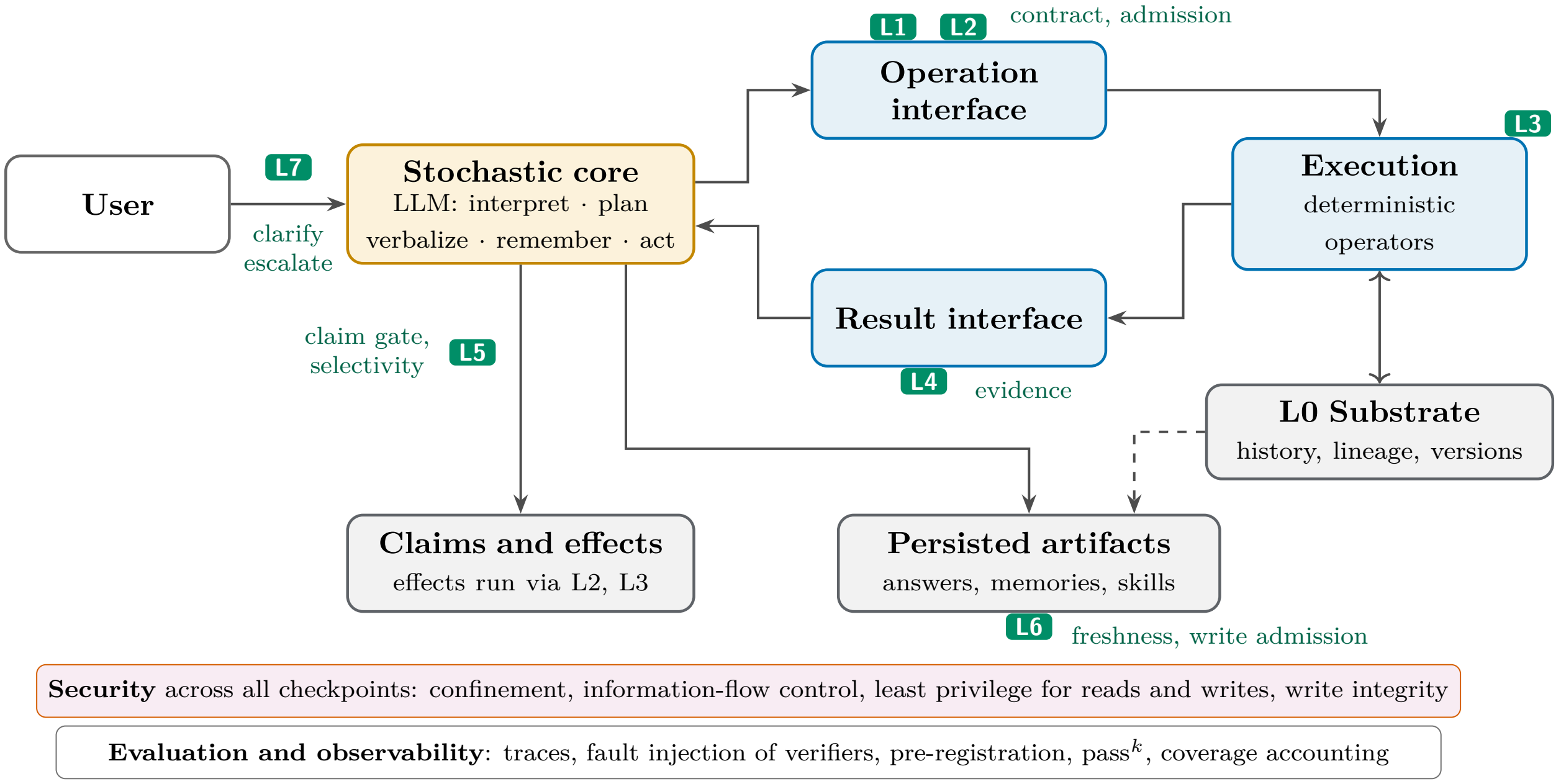

\Cref{fig:layers} places eight protection mechanisms, \lay{0}--\lay{7}, on the dataflow of \Cref{fig:model}.
\lay{0} is a substrate that records what the others need; \lay{1}--\lay{7} are checkpoints where obligations are
discharged or surfaced. They are not a strict stack: each checkpoint depends on what the substrate and earlier
checkpoints recorded, and several act on the same artifact at different times. The model is a \emph{stochastic core}
in the middle. Model-side research raises the probability that it discharges the obligations it keeps
(interpretation, planning, verbalization, calibrated confidence, stability under pressure); the checkpoints bound
what happens when it does not. \Cref{tab:layers} summarizes guarantees, assumptions and costs.

\paragraph{\lay{0} Substrate: history, lineage, versions, and what enters.}
This checkpoint supports execution correctness and temporal validity. A store that overwrites in place cannot answer ``what did we believe on March~1'' and
cannot re-decide a saved answer's validity. The mechanisms are classical: valid and transaction
time~\cite{snodgrass1985taxonomy,kulkarni2012temporal}, append-only logs, lineage and
provenance~\cite{buneman2001where,green2007provenance}. They separate
\emph{update} (the world changed) from \emph{revision} (we were wrong), the distinction belief-revision theory draws
between Katsuno--Mendelzon update and AGM revision~\cite{katsuno1991propositional,alchourron1985logic}. Agent memory
systems are converging on bi-temporal storage~\cite{rasmussen2025zep,wang2026less}; in one preprint a
deterministic supersession rule drove served-stale answers to about zero~\cite{yadav2026temporal}. The substrate should
also record the \emph{non-data} inputs of each computation (model identifier and version, prompts, tool versions, the
environment a skill was validated in) and should validate what enters: constraints, data validation and entity
resolution~\cite{schelter2018automating,breck2019data}, with model-extracted facts marked as such. The substrate is written by the system, not by the agent: agents can be
induced to delete their own traces~\cite{qin2026llm}, so history and audit records must be append-only and
tamper-evident, outside the agent's write path~\cite{ardebili2026novafabric}. Lakehouse time travel and dataset version
control~\cite{armbrust2020delta,maddox2016decibel} provide transaction-time
versions but not valid time; bi-temporal graph stores are the closer research peers~\cite{rost2021bitemporal,hou2024aeong}. \emph{Guarantee}: by
construction, past belief states are reconstructible. \emph{Weakens for agents}: when facts are extracted from text,
the timeline itself is uncertain. Retention and legal erasure also conflict with ``history is never destroyed''; a
substrate must support deletion that is itself recorded, so that dependent artifacts become \textsc{undecidable}
rather than silently wrong.

\paragraph{\lay{1} Interface contract: what may be referenced and what comes back.}
The interface contract supports admissibility and moves several checkable responsibilities out of the model. The operation interface exposes typed
operators with declared argument \emph{and result} schemas, a reference language without free arithmetic, and
identifiers only from the user's input or a resolver. Checking declared output fields can convert silent runtime failures into repairable rejections. Typed semantic
operators~\cite{patel2024semantic,liu2024declarative}, typed runtimes for data analysis~\cite{yan2026fail} and semantic
layers that declare measures and units~\cite{rumiantsau2026semantic} are other instances. Constrained
decoding~\cite{willard2023efficient,dong2024xgrammar} guarantees syntax, not meaning.
\emph{Guarantee}: by construction, for what the language cannot express. \emph{Cost}: expressiveness; a fixed algebra limits which questions can be asked at all.

\paragraph{\lay{2} Admission: static verification, authorization, cost.}
Admission supports admissibility and effect safety. Before anything runs, a verifier checks the whole plan (types, references, temporal
arguments, grounding, declared output fields, privileges, estimated cost) and returns \emph{repairable} rejections that
name a legal alternative. Whole-plan checking sees dataflow and cost before execution, which per-call schemas in
ReAct-style loops~\cite{yao2023react} cannot. Agent-security work converges on the same pattern: enforce policy
deterministically, outside the model~\cite{debenedetti2025defeating,wang2025agentspec}.
\emph{Guarantee}: verified, relative to the policy and cost model. \emph{Weakens for agents}: cost models tuned to single
operators may not carry over to multi-step plans, and an underestimate is the unsafe direction. Policies proposed by an LLM put a stochastic component in the trust base unless only
narrowing changes are automatic~\cite{shi2025progent}.

\paragraph{\lay{3} Execution: deterministic, traced, transactional, bounded.}
Execution supports correctness and effect safety. Operators execute deterministically over a pinned basis and emit content-addressed
traces. Because inference is not reproducible by default on common serving stacks, even at temperature
zero~\cite{atil2024non}, and API users cannot make it so although batch-invariant kernels can~\cite{he2025defeating},
audit must rest on \emph{logged} model outputs, treated as external inputs; deterministic replay applies to the data
system. For
effects, the classical machinery is transactions, sagas and
compensation~\cite{haerder1983principles,garciamolina1987sagas}; agent runtimes add effect outboxes, shadow state,
buffered release and progress-aware commit~\cite{mohammadi2026atomix,chen2026cordon}.
Agents also explore by branching: copy-on-write branches and cross-store branching make speculative execution cheap
and isolated~\cite{zhou2026chronos,ang2026branchbench}, and new commit criteria require that what the
model saw and the policy it relied on are unchanged at commit~\cite{he2026ai,peng2026stateful}. Ordinary fault
tolerance belongs here too: idempotency keys for retried calls, timeouts, and runtime budgets that
abort loops as refusals rather than as partial answers. \emph{Guarantee}: by construction for deterministic operators;
verified for effect release. \emph{Weakens for agents}: a transaction body written by an agent cannot be replayed, so
redo-based recovery is unavailable and only undo, compensation and confinement remain~\cite{mraz2026tracking};
isolation can leak through the context window, which no lock covers; irreversible effects make atomicity a property of
how tools are classified; and undo fails mostly at identifying which state to restore~\cite{sah2026evoundo}.

\paragraph{\lay{4} Evidence: results that carry their conditions.}
The evidence checkpoint supports execution correctness and makes the conditions of a result explicit. Each result carries a descriptor of its basis, execution and delivery completeness,
exact cardinality when known, approximation and provenance; conditions propagate through composition, so a count over
a filtered page cannot be certified at a later step whose own metadata looks complete. For
retrieval, the analogous descriptor is a recall statement and the snapshot or consistency level the read actually
observed; per-query recall guarantees~\cite{loh2026per} and declarative recall
targets~\cite{chatzakis2025darth} for approximate indexes are beginning to appear, but they bound index loss, not whether the right scope was searched, and
vector stores expose consistency as a query-time knob rather than a returned fact~\cite{guo2022manu}. Claim-level
provenance mechanisms verify an aggregate claim by a semantic query with semiring provenance~\cite{lee2026evergreen} or
find an input subset that reproduces an LLM operator's answer~\cite{lin2026bolt}; both verify against the retrieved
set and do not establish that the set was complete. Classical
precedents bound what is possible: provenance semirings compose lineage~\cite{green2007provenance}; authenticated data
structures and verifiable SQL prove soundness \emph{and completeness} against a committed database even with an
untrusted server~\cite{pang2005verifying,zhang2017vsql}; approximate query processing attaches
error bounds~\cite{hellerstein1997online,agarwal2013blinkdb}; semantic-operator cascades attach precision and recall
targets relative to an oracle~\cite{kang2020approximate,zeighami2025cut}. \emph{Guarantee}: verified, relative to
backend semantics and adapter truthfulness, or cryptographic. \emph{Cost}: recording conditions is cheap, but establishing some of them is not: an exact count may require
scanning past the page, and in general evidence can cost as much as the query it certifies.

\Cref{tab:descriptor} makes the descriptor concrete for the relational case. A descriptor for a result $S$ records the
delivered rows $R\subseteq S$, a cardinality fact (exact $|S|=n$, a lower bound, or nothing), whether $S$ is ordered and
by what key, an approximation bound if any, and the basis $\beta$. Propagation rules give the descriptor of an
operator's output from the descriptors of its inputs; they compute the certain part and an interval for aggregates, in
the range semantics of \Cref{sec:formal}. The rules apply equally when the agent writes free-form SQL rather than calling
typed operators: the DBMS knows whether a \texttt{LIMIT}, a page size or a tool-side cap truncated the result and what
snapshot was read, so it can emit the base descriptor, and the rules then follow the query's algebra. Open SQL makes
\lay{1} weaker (more can be expressed) but leaves \lay{4} intact; what it loses is that the system can no longer check
declared result fields before execution.

\begin{table}[!htbp]
\centering\footnotesize
\caption{A result descriptor and propagation rules (sketch). Inputs have delivered rows $R$ ($R_1,R_2$) out of full
results with $n$ ($n_1,n_2$) rows; $m=n-|R|$ rows are undelivered ($m$ is unbounded when no total is reported). ``Certain'' is what holds in every state of
$\mathrm{Rep}$; intervals are range answers.}
\label{tab:descriptor}
\begin{tabularx}{\textwidth}{@{}L{2.6cm} Y Y@{}}
\toprule
\textbf{Operation} & \textbf{Certain part and completeness} & \textbf{Aggregates and bounds} \\
\midrule
base read & $R$ certain; complete iff $|R|=n$ & $|S|=n$ if exact total reported, else $|S|\ge|R|$ \\
filter $\sigma_p$ & $\sigma_p(R)$ certain; complete iff input complete & count $\in[\,|\sigma_p(R)|,\ |\sigma_p(R)|+m\,]$ \\
projection $\pi$ (set) & $\pi(R)$ certain; complete iff input complete & distinct count $\in[\,|\pi(R)|,\ |\pi(R)|+m\,]$ \\
join $\bowtie$ & $R_1\bowtie R_2$ certain; complete iff both inputs complete & no useful upper bound in general; $\le n_1$ under a key--foreign-key constraint \\
aggregate $\gamma$ & value certain only if input complete & $\mathrm{count}=n$ if an exact total is reported, else $\ge|R|$; $\max\ge\max(R)$, $\min\le\min(R)$; $\mathrm{sum}\ge\mathrm{sum}(R)$ for non-negative values, $\le \mathrm{sum}(R)+m\cdot v_{\max}$ if a value bound $v_{\max}$ is declared \\
top-$k$ over a prefix & certain iff $R$ is a prefix of $S$ ordered by the same key and $k\le|R|$ & on an unordered page: only ``at least'' statements \\
approximate input & certain part as above, widened by the declared error & absolute error bounds add under sums; probabilistic bounds need the failure probabilities combined (\Cref{sec:compose}) \\
any & basis of output = basis of inputs; mixed bases are flagged & provenance composes as in provenance semirings~\cite{green2007provenance} \\
\bottomrule
\end{tabularx}
\end{table}

\paragraph{\lay{5} Claim gate and selectivity: what may be said, and whether to say it.}
The claim gate connects result evidence to what the system is willing to state or release. The report is decomposed into typed claims citing the results they rest on; a verifier
re-checks each claim by certain-answer reasoning over the cited descriptor (\Cref{sec:formal}), and unsupported claims
are gated. The binding between text and claims must itself be checkable: in the style of proof-carrying
numbers~\cite{solatorio2025proof}, the text shown to the user is rendered from claim objects, so every number or entity
in it occupies a slot bound to a checked claim, and free prose that makes no typed claim is shown as uncertified rather
than gated or silently trusted. A claim gate therefore trades some coverage for a stronger release condition. Attribution methods for generated text judge support
probabilistically~\cite{rashkin2021measuring,gao2023enabling}; where the evidence is a query result, support
can be checked deterministically. For what remains uncertain, mainly interpretation, selectivity decides
whether to answer: confidence as a control signal~\cite{li2026confidencecontrol}; thresholds with a risk guarantee,
which need labeled calibration data from the deployment distribution, not merely scores that look
calibrated~\cite{angelopoulos2021learn,li2026calibrationrl}; conformal factuality and risk
control~\cite{mohri2024language,angelopoulos2024conformal}; selective
prediction~\cite{geifman2017selective,wen2025know}, and accept-or-escalate cascades whose savings
depend on the cheap stage being competent where it is confident~\cite{li2026jev}. Evaluation that scores accuracy
alone rewards guessing over abstaining~\cite{kalai2025language}, which is why coverage must be reported with accuracy.
Gating is selective prediction with an
exact special case: for claims checkable against result rows, the tolerated risk is zero and no threshold is needed.
Every statistical guarantee should name its oracle of record, risk level and calibration-set
version~\cite{liu2026certified}. \emph{Guarantee}: verified for gated claim types; statistical for selectivity, under an
exchangeability assumption that data change and retroactive correction break~\cite{tibshirani2019conformal}, so
calibration sets are themselves derived artifacts for \lay{6}. \emph{Cost}: coverage.

\paragraph{\lay{6} Lifecycle: freshness and admission of persistent state.}
Lifecycle checks protect persisted state from silent staleness and unsupported updates. Incremental view maintenance keeps definable views equal to
recomputation~\cite{gupta1993maintaining,budiu2023dbsp}; truth maintenance retracts conclusions when
premises are retracted~\cite{doyle1979truth,dekleer1986assumption}. Neither transfers exactly: a model-written summary has
no delta rule, and its dependence on parametric knowledge is not recorded. A \emph{one-sided} contract does transfer:
record a conservative superset of each artifact's inputs, including model and environment versions, and answer
\textsc{fresh}, \textsc{possibly stale} or \textsc{undecidable}, never calling a changed artifact fresh. The price of one-sidedness is needless refresh when lineage is loose, so lineage precision decides whether
incremental refresh beats recomputation. Incremental maintenance of LLM-augmented tables already records input versions per
refresh~\cite{yadav2026enzyme}, and transactional agent memories commit writes with evidence and
cascade retractions to derived records~\cite{li2026memtx,wang2026toki}; extending lineage to model, tool and
environment versions with one-sided verdicts is the remaining step. The same checkpoint governs writes into
persistent state: memory and skill writes should carry a source and pass admission, versioned with
rollback~\cite{su2026chronomem,li2026dynamicskills}, because memories can be poisoned through ordinary
queries~\cite{dong2025memory}. Write admission is also how this checkpoint serves \ob{8}: a belief held in memory
changes only through an admitted write with a source, so conversational pressure alone cannot revise it.
\emph{Guarantee}: verified, one-sided. \emph{Cost}: needless refreshes when lineage is
loose.

\paragraph{\lay{7} Interaction and governance.}
Interaction and governance address the part of reliability that cannot be reduced to deterministic execution checks. Clarification is a query-processing decision: ask when the expected
reduction in error or work exceeds the interaction cost~\cite{horvitz1999principles,kuhn2022clam}.
Assumptions the agent did make should be recorded and shown with the answer as an \emph{interpretation record} that can
be confirmed, reused and audited. Beliefs about data are read from the substrate and change through writes that carry a
source, not through conversational pressure~\cite{li2026epistemic}. Escalation routes uncertain cases to stronger models
or people~\cite{li2026jev}. Human oversight has its own failure modes: automation is misused and
disused~\cite{parasuraman1997humans}, supervising highly automated systems is ironically
hard~\cite{bainbridge1983ironies}, analysts miss plausible-looking errors in AI-assisted analyses~\cite{gu2023analysts},
and forcing functions reduce overreliance at a usability cost~\cite{bucinca2021trust}. \emph{Guarantee}: human or
monitored.

\paragraph{Cross-cutting: security and evaluation.}
Confinement, information-flow control and least privilege apply at every checkpoint: control and data flow derived
from the trusted request~\cite{debenedetti2025defeating}, labels that constrain sinks~\cite{costa2025securing},
narrowing-only privilege updates~\cite{shi2025progent}. The utility cost is real: one provably secure design solves 77\%
of AgentDojo tasks against 84\% undefended~\cite{debenedetti2025defeating}. Reads need least privilege as much as writes, since an answer can leak: data flow
control enforces tuple-level release policies inside the DBMS through provenance rewriting~\cite{summers2026data}, while
benchmarks show that models asked to respect access policies in text-to-SQL both violate and over-refuse
them~\cite{fei2026benchmarking}, and MCP database servers inherit SQL injection and over-privilege~\cite{padilla2026exposed}. Evaluation is the other cross-cutting concern: each guarantee should be tested by fault
injection against an oracle with deliberate negative cases, agreement should be asserted on the \emph{shape} of what was
compared (non-empty answers, corrections present, truncation exercised){}, traces should make
failure attribution mechanical, since models diagnose failed traces poorly~\cite{deshpande2025trail}, and evaluation
criteria need their own revision loop~\cite{shankar2024who,shankar2024spade}.

\begin{table}[!htbp]
\centering\footnotesize
\caption{Checkpoints, mechanisms, guarantees and limits. ``Weakens'' names the classical assumption that fails when
the author of operations is an LLM.}
\label{tab:layers}
\begin{tabularx}{\textwidth}{@{}L{1.75cm} L{3.3cm} L{2.2cm} Y L{2.4cm}@{}}
\toprule
\textbf{Stage} & \textbf{Mechanisms} & \textbf{Guarantee} & \textbf{Weakens for agents} & \textbf{Cost} \\
\midrule
\lay{0} Substrate & valid/transaction time; append-only log; lineage; model and environment versions; input validation & by construction & extracted timelines are uncertain; erasure must be recorded & storage; query complexity \\
\lay{1} Interface & typed operators; result schemas; grounded identifiers; semantic layer & by construction & user concepts absent from schemas; syntax $\neq$ meaning & expressiveness \\
\lay{2} Admission & static plan verifier; privileges; cost guard; repair payloads & verified (policy-relative) & cost models do not compose to plans; LLM-authored policies & false rejections; repair loops \\
\lay{3} Execution & deterministic operators; traces; outbox; compensation; idempotency; budgets & by construction; verified release & no redo; isolation leaks via context; irreversible effects & latency; interposition \\
\lay{4} Evidence & completeness, cardinality, basis, approximation, provenance; recall for retrieval; certificates & verified (trust-relative); cryptographic & oracle-relative statistics; adapters in the trust base & up to the cost of the query \\
\lay{5} Claim gate & typed claims; certain-answer check; risk-controlled selectivity & verified; statistical & calibration not portable; exchangeability breaks over time & coverage \\
\lay{6} Lifecycle & conservative lineage; one-sided freshness; versioned memory and skills; write admission & verified, one-sided & no delta rules for model outputs; poisoning via legitimate channels & needless refreshes \\
\lay{7} Governance & clarification operator; interpretation records; escalation; audit & human; monitored & intent unobservable; overreliance & user burden \\
\bottomrule
\end{tabularx}
\end{table}

\subsection{Composing the checkpoints}
\label{sec:compose}

\paragraph{Containment, not probability.}
Checkpoints \lay{1}--\lay{5} bound the \emph{consequences} of a wrong plan: an inadmissible plan does not run, an
unsupported claim is not released, a stale answer is not served as fresh. They do not lower the probability that the
model chose the wrong plan, which is why an otherwise well-protected system can still fail on the plan itself. Only \lay{7} and better models address that probability.

\paragraph{What composes and what does not.}
If each checkpoint is sound in one direction (it may refuse wrongly but never certifies wrongly), their conjunction is
sound in the same direction, but only relative to the union of their trust bases: the claim gate is sound only if
the descriptors it reads are truthful. Statistical guarantees compose less simply. Write a pipeline as
$V \circ E \circ P$: an LLM planner $P$, execution $E$ over a stated snapshot, and an LLM verbalizer $V$. Relative to
trusted operators, $E$ adds no error, and claims of $V$ checkable against $E$'s result add none after gating. The
residual risk lies in $P$ (interpretation) and in the claims of $V$ that cannot be checked. If each has a marginal
guarantee, $\Pr[\text{error}_P]\le\alpha_P$ and $\Pr[\text{error}_V]\le\alpha_V$, a union bound gives
$\Pr[\text{error}]\le\alpha_P+\alpha_V$ whatever the dependence between the two; correlation between LLM components
(\rc{9}) makes the bound loose, not invalid. Two things do not compose this way. First, selection: a guarantee over all
requests does not transfer to released ones, since $\Pr[\text{error}\mid\text{released}]\le\alpha/\Pr[\text{released}]$,
which is weak when coverage is low; a release policy needs a guarantee conditional on release, as in selective risk
control~\cite{geifman2017selective,angelopoulos2024conformal}. Second, magnitudes: union bounds combine failure
probabilities, not error sizes; per-operator precision or error targets give a pipeline target only through how each
downstream operator propagates error (\Cref{tab:descriptor}). This is why deterministic, one-sided checks should be used wherever possible, while statistical checks are reserved for the parts that remain genuinely uncertain.

\paragraph{The trust base.}
Each verified guarantee is relative to trusted components: operators, adapters, the verifier, the policy. Making that
base explicit, small and tested is the database analogue of proof-carrying code, where a consumer checks a proof
instead of trusting a producer~\cite{necula1997pcc}, and of result checking, where a program's output is checked by a
simpler program~\cite{blum1995checking}.

\paragraph{Coverage accounting.}
Every checkpoint that refuses buys soundness with coverage. Evaluation should therefore report the share of requests answered, accuracy on answered requests, and the rate of unsupported claims shown to users, alongside the same measures when the guardrail under study is disabled.

\section{Open Problems}
\label{sec:open}

The architecture above is intentionally incomplete. It identifies where a system can place guardrails, but several obligations cannot yet be discharged cleanly or cheaply. We see six research directions that are especially important for database systems.

\paragraph{Make intent explicit enough to check.}
Execution correctness does not establish that the executed operation matches the user's need. Analytical SQL can be executable yet semantically wrong, and agentic text-to-SQL systems can silently commit to assumptions that were never confirmed~\cite{mahmud2026exploring,li2026beyond}. A useful systems abstraction is an \emph{interpretation record}: a typed description of assumptions about metric, population, grain, time, nulls, duplicates, and tie-breaking that is attached to the plan and result. The open problem is to decide which assumptions can be extracted from the plan, which must be elicited from the user, and when clarification is worth its cost. One promising direction is to execute plausible interpretations cheaply and ask only when they produce materially different answers.

\paragraph{Treat evidence as a composable data object.}
Agent-facing results increasingly mix exact relational operators, retrieval, approximate query processing, and semantic operators. Each stage may know something different about completeness, approximation error, recall, provenance, or staleness, but those conditions are rarely represented in a form that composes across the pipeline. Database provenance and incomplete-information research provide useful foundations~\cite{green2007provenance,libkin2016sqls}, while approximate and semantic operators provide local bounds~\cite{hellerstein1997online,zeighami2025cut}. The challenge is to define an evidence algebra whose annotations propagate through operators and can be consumed by a claim gate. This also creates a new optimization problem: choose plans not only for latency or cost, but for the evidence needed by the claims the system expects to release.

\paragraph{Define a checkable boundary between results and claims.}
A value can be correct while a sentence about that value is unsupported. Reliable systems therefore need a restricted claim language or another representation in which support can be checked against returned evidence. The difficult cases are not only factual extraction but quantifiers, comparisons, causal wording, and claims over incomplete sets. Work on attribution and grounded generation shows how to measure support probabilistically~\cite{rashkin2021measuring,min2023factscore}; database systems can go further when claims refer to structured query results by checking some claim forms deterministically. The research question is how expressive such a claim language can become before verification becomes as hard and opaque as generation itself.

\paragraph{Make persisted agent state freshness-aware.}
Answers, memories, skills, indexes, thresholds, and calibrators are all derived artifacts. They can become stale when data are corrected, tools change, models are updated, or a dependency is reinterpreted. Temporal databases, provenance, and incremental view maintenance already provide pieces of the solution~\cite{snodgrass1985taxonomy,gupta1993maintaining}; recent memory systems are beginning to expose temporal structure as well~\cite{rasmussen2025zep,yadav2026temporal}. What is missing is a common validity contract for heterogeneous artifacts. The system should be able to answer at least one of three questions cheaply: still valid, definitely stale, or unknown. The unknown case must be visible rather than silently treated as fresh.

\paragraph{Give agent actions transactional boundaries.}
Agents do more than query: they write data, call services, send messages, and trigger workflows. Classical transactions and sagas provide atomicity, recovery, isolation, and compensation, but agent execution adds speculative branching, long-running plans, external side effects, and policies that may change between observation and commit~\cite{haerder1983principles,garciamolina1987sagas}. Recent agent runtimes explore buffered effects and commit-time checks~\cite{mohammadi2026atomix,chen2026cordon}. The open problem is to define transaction semantics for an agent whose plan is stochastic and whose steps may span several systems. At minimum, tools should expose effect classes, idempotency properties, and commit or compensation behavior that the runtime can enforce independently of the model.

\paragraph{Build independent assurance and evaluation.}
Every guarantee is relative to an oracle, label source, checker, and trust base. LLM judges can share errors with the systems they evaluate, confidence can be poorly aligned with correctness, and benchmark labels themselves can be wrong~\cite{jin2026pervasive,liu2026certified}. Reliable agentic systems therefore need evaluation that separates \emph{coverage} from correctness and measures the failure of the guardrails themselves. We recommend reporting at least: accuracy on answered requests, abstention or coverage, unsupported claims exposed to users, repeated-run variability, behavior under incomplete-result and stale-state traps, and recovery from duplicated or failed effects. Statistical guarantees should name their oracle of record; system guarantees should name the components they trust. Fault injection and metamorphic testing are particularly useful because they test whether a protection mechanism actually catches the class of failure it claims to catch~\cite{rigger2020testing}.

These directions share one theme: the data system should not try to make the model infallible. It should make important conditions explicit, preserve them long enough to be checked, and prevent claims or effects from crossing a commit boundary when the required conditions are known not to hold. Better models then improve coverage and usability inside a system whose failure envelope is visible.

\section{Conclusion and Outlook}
\label{sec:outlook}

This paper advances a systems perspective on reliability rather than a claim that the nine obligations are the only possible decomposition. Other decompositions are possible, and organizational or process failures can lie outside the system boundary considered here~\cite{sambasivan2021everyone}. The root causes likewise identify recurring fault sources and design choices rather than mutually exclusive causal categories. Their purpose is to connect different kinds of failure to different system responses. Our literature grounding is broad but deliberately selective: we use prior work to support particular observations, mechanisms, and open questions rather than to claim exhaustive coverage of a rapidly changing area.

The larger point is that agentic data systems change a basic assumption behind traditional data interfaces. The client that formulates an operation, interprets its result, remembers it, and acts on it may itself be stochastic. Better models will reduce some failures, but they cannot recover conditions that the data system never exposed, preserve history that the system discarded, or turn an unsupported claim into a supported one after the fact. Reliability therefore cannot rest on model capability alone.

We have argued for treating reliability as an end-to-end systems contract. The obligations identify what can go wrong between user intent and system effects; the root causes explain why those failures arise; and the guardrails place checks where the relevant information already exists. This view suggests a shift in emphasis for database systems: from asking only whether an operation executed correctly to asking what assumptions, evidence, validity conditions, and effect boundaries must accompany the result so that an agent can use it safely.

The goal is not an infallible agent. It is a data system in which unsupported claims are difficult to release silently, stale artifacts are recognizable before reuse, side effects have clear boundaries, and unresolved uncertainty remains visible to whoever must act on it. As agentic interfaces become a normal way to access and modify data, these properties offer a concrete agenda for the database community: make reliability conditions explicit, preserve them across the agent--system boundary, and enforce them wherever the system has enough information to do so.

\section*{Acknowledgments and use of AI tools}

LLM tools assisted literature discovery, preliminary screening, organization of research notes, reference checking, and language editing under the author's direction.  The author developed the paper's framing and arguments, reviewed the retained evidence and references, and remains responsible for the content and conclusions. LLM-assisted screening was used as a discovery and organization aid rather than as evidence of exhaustive coverage or as the basis for quantitative claims about the literature.

\bibliographystyle{plainnat}
\bibliography{refs}

\end{document}